\documentclass[preprint,nofootinbib]{revtex4}
\usepackage{graphicx}
\usepackage{color}
\def\w{{\rm w}}

\def\tev{~{\rm TeV}}

\def\mh{m_H^{}}
\def\yt{y_T^{}}
\def\ywl{y_{W_L}^{}}
\def\ywh{y_{W_H}^{}}
\def\yp{y_{\Phi^+}^{}}
\def\ypp{y_{\Phi^{++}}^{}}
\def\ygf{y_{G_F}^{}}

\def\lsim{\mathrel{\raise.3ex\hbox{$<$\kern-.75em\lower1ex\hbox{$\sim$}}}}
\def\gsim{\mathrel{\raise.3ex\hbox{$>$\kern-.75em\lower1ex\hbox{$\sim$}}}}

\begin{document}

\bibliographystyle{revtex}

\preprint{
 {\vbox{
 \hbox{MADPH--03--1319}
 \hbox{hep-ph/0302188}}}}

\vspace*{2cm}

\title{Loop induced decays of the Little Higgs: $H\to gg,\ \gamma\gamma$}

\author{Tao Han, Heather E. Logan, Bob McElrath, and Lian-Tao
Wang\footnote{han@pheno.physics.wisc.edu\\ logan@pheno.physics.wisc.edu\\
mcelrath@pheno.physics.wisc.edu\\ liantaow@pheno.physics.wisc.edu}}
\affiliation{\vspace*{0.1in}
Department of Physics, University of Wisconsin, 1150 University
Avenue, Madison, WI 53706}

\vspace*{1.0cm}

\begin{abstract}
We analyze the loop induced decays of the Higgs boson into pairs of
gluons and photons in the Littlest Higgs model.  
We find that the deviation of the partial widths for these decays
relative to their Standard Model values scales with $1/f^2$, where
$f \sim {\rm TeV}$ is the mass scale of the new heavy particles
in the model.
For $f = 1$ TeV, $\Gamma(H \to gg)$ is reduced by $6-10\%$ and
$\Gamma(H \to \gamma\gamma)$ is reduced by $5-7\%$ compared
to their Standard Model values.  
While the LHC and a linear $e^+e^-$ collider would be sensitive 
to these deviations
only for relatively low values of $f \lsim 650$ GeV, a photon collider could 
probe the deviation in $\Gamma(H \to \gamma\gamma)$ up to 
$f \lsim 1.1$ (0.7) TeV at the 2 (5) $\sigma$ level.
\end{abstract}


\maketitle

\section{Introduction}

The Standard Model (SM) of the strong and electroweak interactions has
passed stringent tests up to the highest energies accessible today.
The precision electroweak data \cite{Hagiwara:fs} prefer the existence of
a light Higgs boson of mass $\mh\lsim 204$ GeV at $95\%$ C.L. 
The Standard Model 
with a light Higgs boson can be viewed as an effective theory 
valid to a much higher energy scale $\Lambda$, possibly all the way
to the Planck scale. However, without
protection by a symmetry, the Higgs mass is quadratically
sensitive to quantum corrections, rendering the theory with 
$\mh \ll \Lambda$ rather unnatural.

Little Higgs models \cite{littleh,littlesthiggs} 
revive an old idea to keep the 
Higgs boson naturally light.  The key idea is to make the Higgs particle a
pseudo-Goldstone boson of some broken global symmetry.  
Electroweak symmetry breaking is then triggered by a
Coleman-Weinberg \cite{coleman-weinberg} potential, generated by
integrating out heavy degrees of freedom.
The Higgs boson acquires a mass at the electroweak 
scale possibly by radiative corrections. 
The Littlest Higgs (LH) model \cite{littlesthiggs} is a minimal model of this
type.  It contains the minimal matter content necessary to accomplish
the goal of canceling the Higgs mass quadratic divergence to one loop order.
It consists of an $SU(5)$ non-linear $\sigma$-model which
is broken down to $SO(5)$ by a vacuum condensate $f$.
The gauged subgroup $[SU(2)\times U(1)]^2$ is broken at the
same time to its diagonal subgroup $SU(2)\times U(1)$ identified
as the SM electroweak gauge group. A vector-like ``top quark'' is also needed. 
Particles of the {\it same} spin cancel quadratic divergences among themselves.

Due to the symmetry breaking by the vacuum
condensate, the theory has a natural cutoff $\Lambda\approx 4\pi f$ that is
$\mathcal{O}(10 \tev)$. The new heavy states, such as the new gauge bosons,
heavy vector-like quark, and scalars all have masses of the order $f$,
naturally in the TeV range. The effects due to the new
states at low energies have been studied recently and generic constraints
on the model parameters, in particular $f$, are 
obtained \cite{ewdata,JoA,us}
based on the precision electroweak measurements.
Signatures for the new states at future colliders have also been
examined \cite{Burdman,JoA,us}.

In this paper, we study the loop-induced decays $H \rightarrow g g$ and
$H\rightarrow \gamma \gamma$  in the Littlest Higgs model.  There are several
motivations to perform this study. First, while the Higgs boson is 
the central issue of the little Higgs models, its properties have 
not been carefully studied to compare with
the SM expectations. By construction, the Higgs boson interactions are
necessarily extended and modified beyond the SM.
Second, although the new states in the LH model could
be too heavy to be copiously produced at the LHC and quite possibly beyond
the direct reach of a linear collider, quantum corrections due to these states 
running in the loops may reveal an earlier signature.
Third, these loop-induced decays depend primarily on the couplings 
of the Higgs boson to the heavy quarks and gauge bosons, which are 
fixed and characteristic in the LH model.
This is in contrast to the Higgs boson couplings to the light fermions,
which could in principle be generated by some higher-dimensional 
operators, leading to model dependent uncertainties.
Finally, both of these processes are finite at the
one-loop order and therefore independent of the detailed physics
at the cutoff scale $\Lambda$, leading to robust predictions.

It is well known that in the Standard Model, the contribution of a
heavy SM particle to the loop amplitude for $H \to gg$ or 
$H \to \gamma\gamma$ approaches a nonzero constant value for particle
masses much heavier than $H$ (for example, this is a reasonably good
approximation for the top quark loop \cite{heavyt,QCDggh}).  
This ``non-decoupling'' behavior
occurs because the masses of the SM particles are generated by their 
coupling to $H$, so that the mass dependence of the coupling cancels
the mass dependence of the loop integral.  The heavy particles in the 
LH model, on the other hand, get their masses from the $f$ condensate,
so that their couplings to $H$ are not proportional to their masses.
Thus the heavy particle contributions to $H \to gg$ and
$H \to \gamma\gamma$ decouple as $f$ grows.  We will show, in fact,
that the deviation of these partial widths from their SM values scales
with $1/f^2$.

In the next section we lay out the masses and couplings 
of the particles in the LH model relevant to our calculation.
In Sec.~\ref{sec:Hdecays} we present our results for $H \to g g$ and 
$H \to \gamma\gamma$, including their dependence
on the additional free parameters in the LH model, and discuss
the expected experimental sensitivity to these loop-induced 
couplings at future colliders.
In Sec.~\ref{sec:disc} we discuss the robustness of our results under
extensions of the LH model and summarize our conclusions.  
Finally, we present some details of the Higgs sector necessary for our 
calculation in the Appendix.

\section{Higgs boson Couplings to new heavy states}

Any colored fermion that couples to the Higgs boson significantly
will contribute to the decay $H\to gg$. Similarly, any charged particle
that couples to the Higgs boson will contribute to $H\to \gamma\gamma$.
Those states in the Littlest Higgs model include the heavy $SU(2)$ gauge
boson $W_H^\pm$, the vector-like quark $T$, and the charged scalars
$\Phi^\pm, \Phi^{\pm\pm}$. Besides the common condensate $f$ as the
most important scale parameter, the mass and couplings for each 
new state depend upon additional dimensionless parameters. 
The mixing between the two gauge groups 
$SU(2)_1$ and $SU(2)_2$, with couplings $g_1$ and $g_2$ respectively,
is parameterized by $c$.  
The mixing between the top quark and
the heavy vector-like quark $T$ is parameterized by $c_t$.
In the Higgs sector, we introduce a parameter $x$ to 
parameterize the ratio of the triplet and
doublet vevs $(v'/v)$. More explicitly, we have
\begin{equation}
      0<  c  = \frac{g_1}{\sqrt{g_1^2+g_2^2}}<1, \qquad
      0<c_t = \frac{\lambda_1}{\sqrt{\lambda_1^2 + \lambda_2^2}}<1, \qquad
       0\le x = \frac{4fv^\prime}{v^2}<1.
\end{equation}
For the parameters, the electroweak data prefers a small 
$c$~\cite{ewdata,JoA}, 
while the
positivity of the heavy Higgs boson mass requires $v^\prime/v < v/4f$,
{\it i.e.}, $x < 1$. 
One should note that $x = 0$ can be
achieved by tuning the contribution of the heavy vector-like quark to the
Coleman-Weinberg potential against the contribution from the heavy gauge
bosons to make the coefficient $\lambda_{h\phi h}$ in
Eq.~(\ref{hphipotential}) vanish.

\begin{figure}
    \resizebox{\textwidth}{!}{
        \rotatebox{270}{\includegraphics{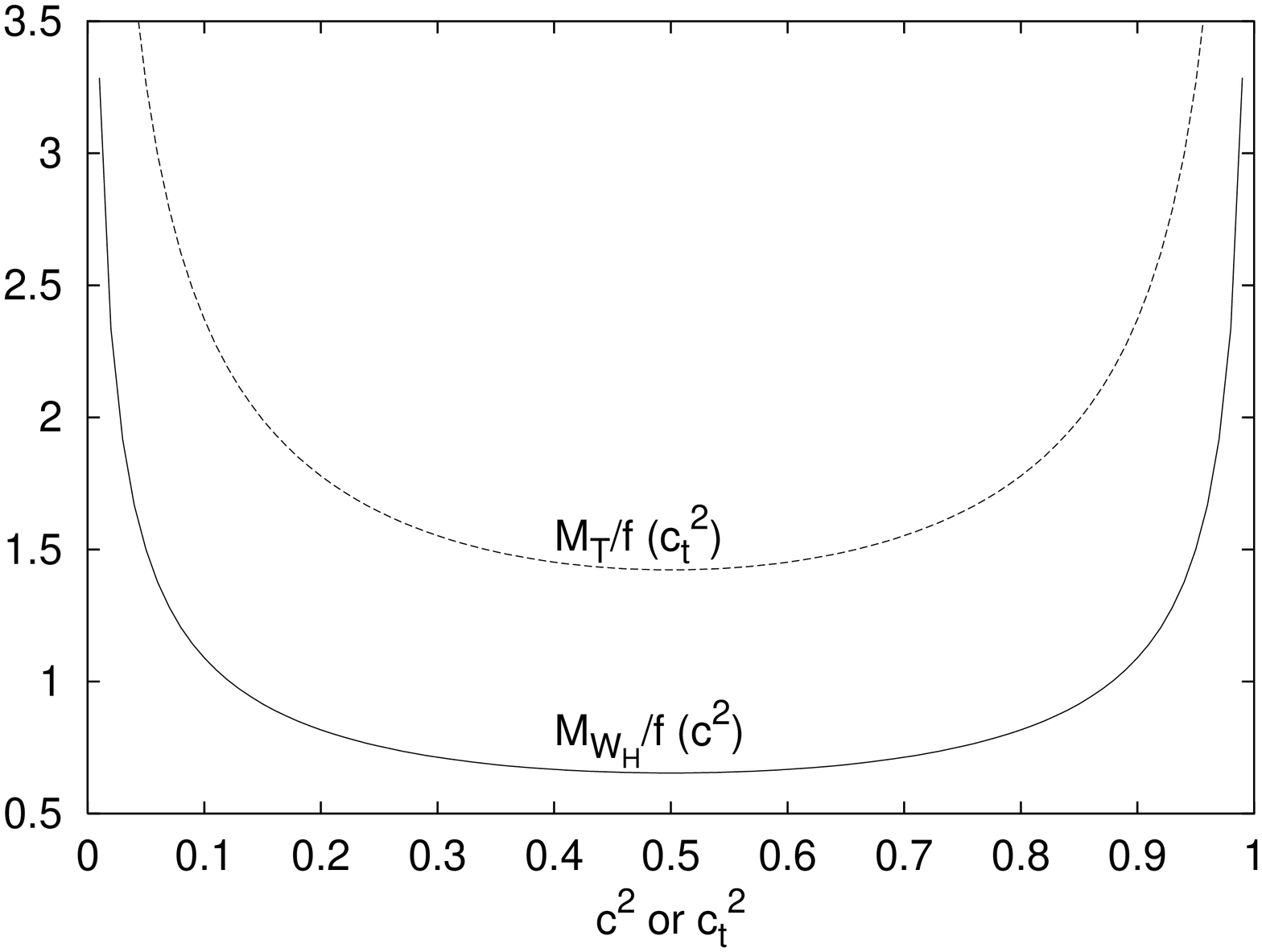}}
        \rotatebox{270}{\includegraphics{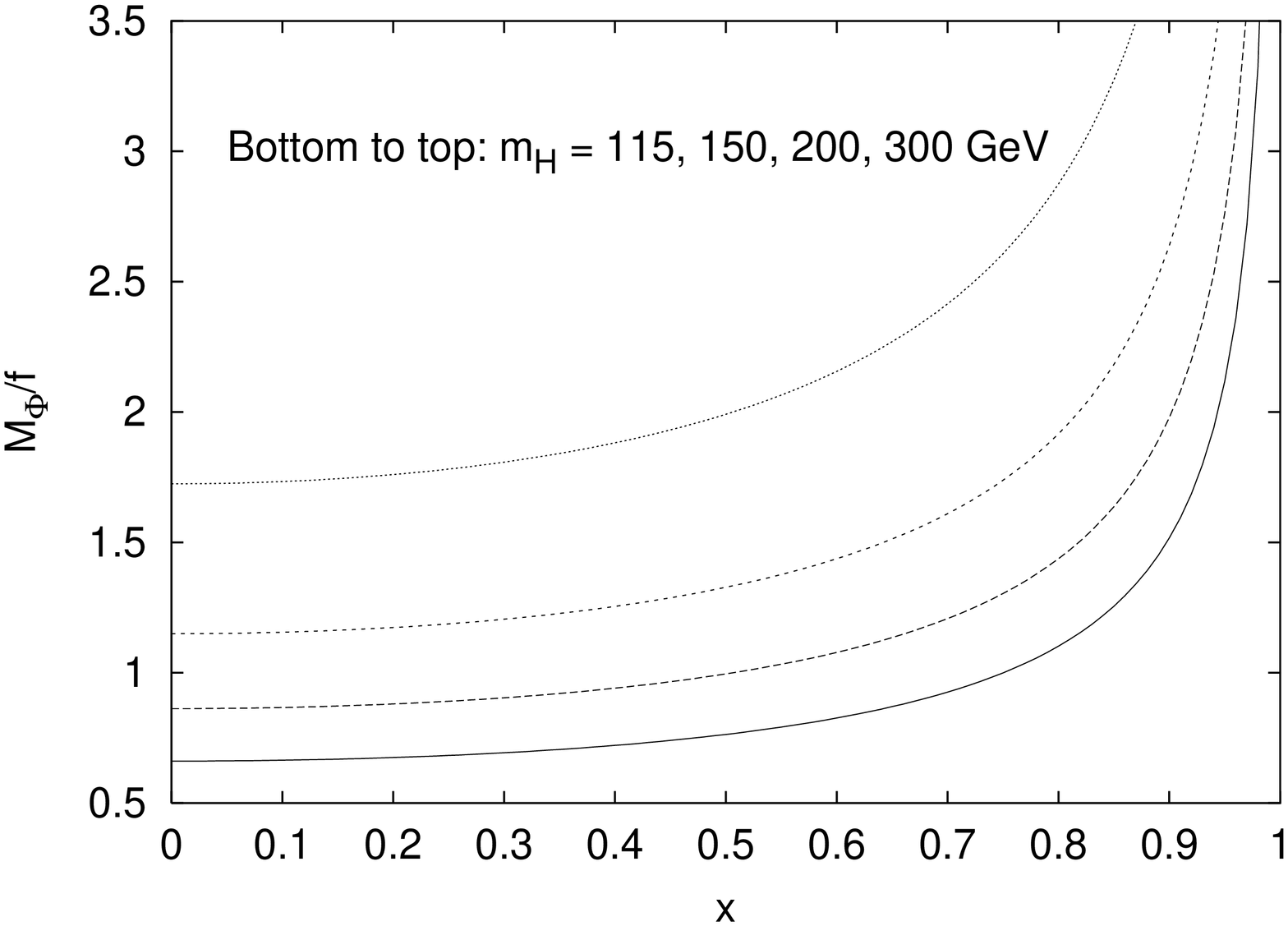}}
    }
    \caption{(left) $M_{W_H}/f$ as a function of $c^2$ and $M_T / f$ as
    a function of $c_t^2$; (right) $M_\Phi/f$ as a function of $x$ for
    various values of the $H$ mass.}
    \label{fig:WHTPhimass}
\end{figure}

The masses of the particles that run in the triangle loop diagrams 
are given to leading order in $v/f$ by
\begin{eqnarray}
	M^2_{W_L}  &=& m_\w^2 \left[ 1 
            - \frac{v^2}{f^2} \left( \frac{1}{6} + \frac{1}{4} (c^2-s^2)^2 
- {x^2\over 4}\right) \right],                  \nonumber \\
    \label{masses}
	M^2_{W_H}  &=& m_\w^2 \frac{f^2}{s^2c^2v^2} ,\quad
	M_T        = f \sqrt{\lambda_1^2 + \lambda_2^2}
            = \frac{ m_t}{s_t c_t } {f\over v} ,\\
	M^2_{\Phi} &=& \frac{2 m_H^2 f^2}{v^2}
                       \frac{1}{[1 - (4 v^{\prime} f/v^2)^2]}
                       = \frac{2 m_H^2}{(1-x^2)}{ f^2\over v^2},   \nonumber
\end{eqnarray}
where $m_\w=gv/2$, $s^2 = 1-c^2$, and $s_t^2 = 1-c_t^2$.
The dependence of these masses on the parameters $c, c_t$ and $x$ for
various representative values of $\mh$ can be seen in 
Fig.~\ref{fig:WHTPhimass}.

In general, all masses that originate with the Standard Model Higgs
mechanism and all couplings to $H$ are modified in the LH model
at order $v^2/f^2$.  We parameterize this by the factors $y_i$,
which are the couplings of $H$ to the particle in the loop normalized
according to the form of the SM Lagrangian as follows \cite{HHG,Martinez}
\begin{eqnarray}
	\mathcal{L} &=& -\frac{m_t}{v} y_t \bar t t H
	- \frac{M_T}{v} \yt \bar T T H
	+ 2 \frac{M_{W_L}^2}{v} \ywl W_L^+ W_L^- H
	+ 2 \frac{M_{W_H}^2}{v} \ywh W_H^+ W_H^- H \nonumber \\
	&& - 2 \frac{M_{\Phi}^2}{v} \yp \Phi^+ \Phi^- H
	- 2 \frac{M_{\Phi}^2}{v} \ypp \Phi^{++} \Phi^{--} H.
\end{eqnarray}
These $y_i$ factors are derived from the Higgs couplings given in 
Appendix A of Ref.~\cite{us}; the Higgs self-coupling factors
$y_{\Phi^+}$ and $y_{\Phi^{++}}$ are derived in the Appendix.  They are
\begin{eqnarray}
    \label{yi}
    \begin{array}{rclrcl}
    y_t           & = & 1 + \frac{v^2}{f^2} \left[ -\frac{2}{3} + \frac{1}{2} x
                        - \frac{1}{4} x^2 + c_t^2 s_t^2 \right]     \qquad &
    \yt           & = & - c_t^2 s_t^2 \frac{v^2}{f^2}                   \\
    \ywl       & = & 1 + \frac{v^2}{f^2} \left[ -\frac{1}{6} 
                        - \frac{1}{4} (c^2-s^2)^2 \right]         \qquad &
    \ywh       & = & - s^2 c^2 \frac{v^2}{f^2} = -M_{W_L}^2/M_{W_H}^2 \\
    \yp    & = & \frac{v^2}{f^2} \left[ -\frac{1}{3} 
                        + \frac{1}{4} x^2 \right]                      \qquad &
    \ypp & = & \frac{v^2}{f^2}
    \mathcal{O}\left( \frac{x^2}{16}\frac{v^2}{f^2}, \frac{1}{16\pi^2} \right).
    \end{array}
\end{eqnarray}
For $\ywl$ we include the correction to the relation between
$M_{W_L}$ and $gv/2$.
One may naively expect some sizable contribution to $H \to \gamma\gamma$
from the doubly-charged
states $\Phi^{\pm\pm}$ due to the $Q^2 = 4$ enhancement of the amplitude.
However, the $H \Phi^{++} \Phi^{--}$ coupling is very small, 
as shown in the Appendix, leading to $y_{\Phi^{++}} \sim v^4/f^4$.
The doubly-charged triplet states thus do not give a significant 
contribution to the amplitude and we will neglect them from now on.

In the SM, the $v$-dependence can be traded for the precisely measured 
$G_F$. In the Littlest Higgs model, however, the relation between $G_F$ 
and $v$ is modified from its SM form, introducing an additional correction
$y_{G_F}$ as 
\begin{equation}
	\frac{1}{v^2} 
	= \sqrt{2} G_F y_{G_F}^2, \qquad {\rm where} \qquad
	y_{G_F}^2 = 1 + \frac{v^2}{f^2} 
	\left[ -\frac{5}{12} + \frac{1}{4} x^2 \right].
\end{equation}
This correction must also be taken into account
when comparing $H$ decay rates in the LH model to the SM predictions
with $G_F$ as input.  Note that $y^2_{G_F} < 1$ for $0 \leq x < 1$,
which tends to suppress the $H$ decay rates.
For a fixed value of $f$, only three parameters of the Littlest Higgs
model affect the loop-induced Higgs partial widths: $c$, $c_t$ and $x$.

\section{\label{sec:Hdecays}$H \to gg$ and $H \to \gamma\gamma$}

\begin{figure}[tb]
    \resizebox{\textwidth}{!}{
        \rotatebox{270}{\includegraphics{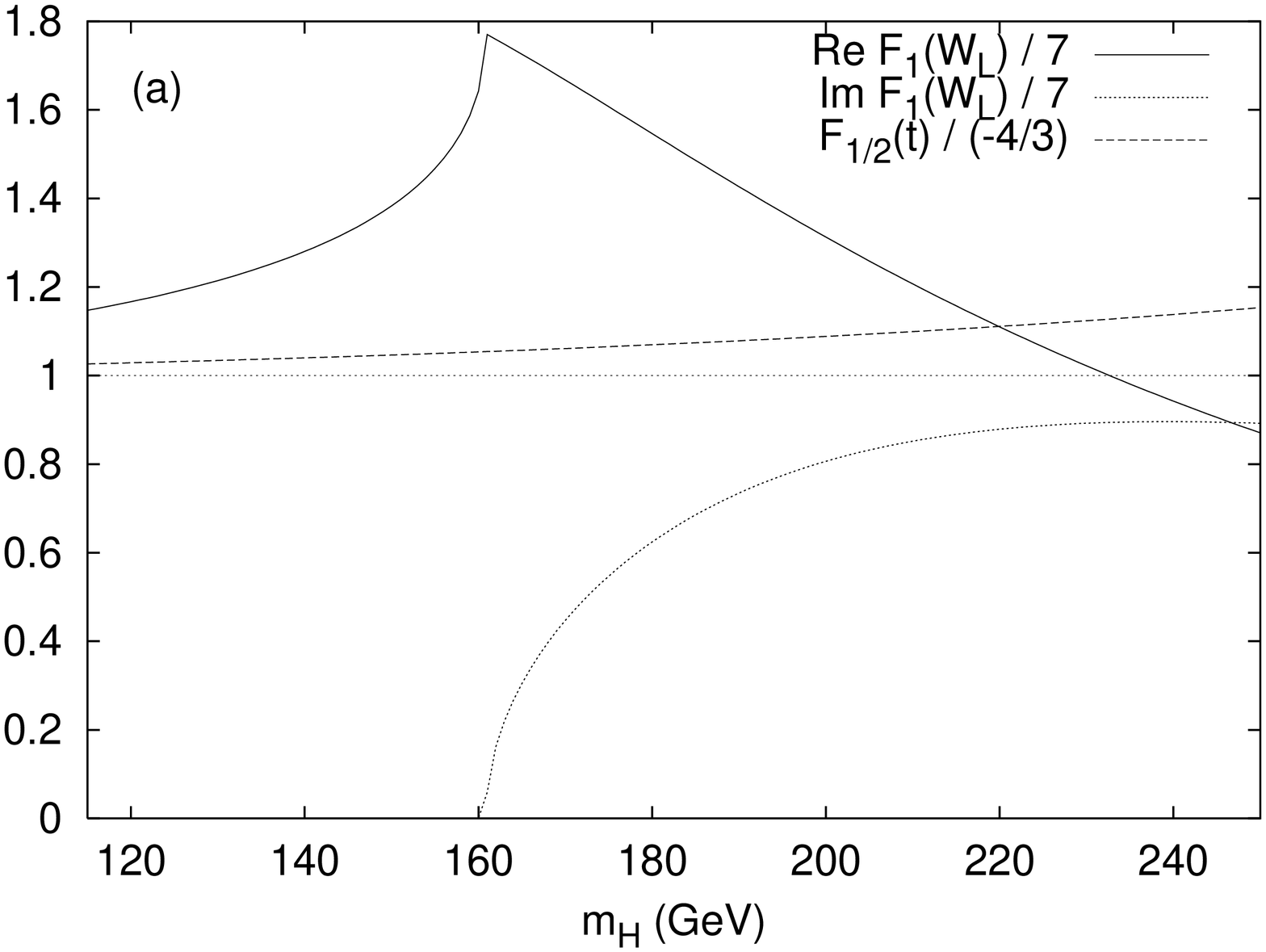}}
        \rotatebox{270}{\includegraphics{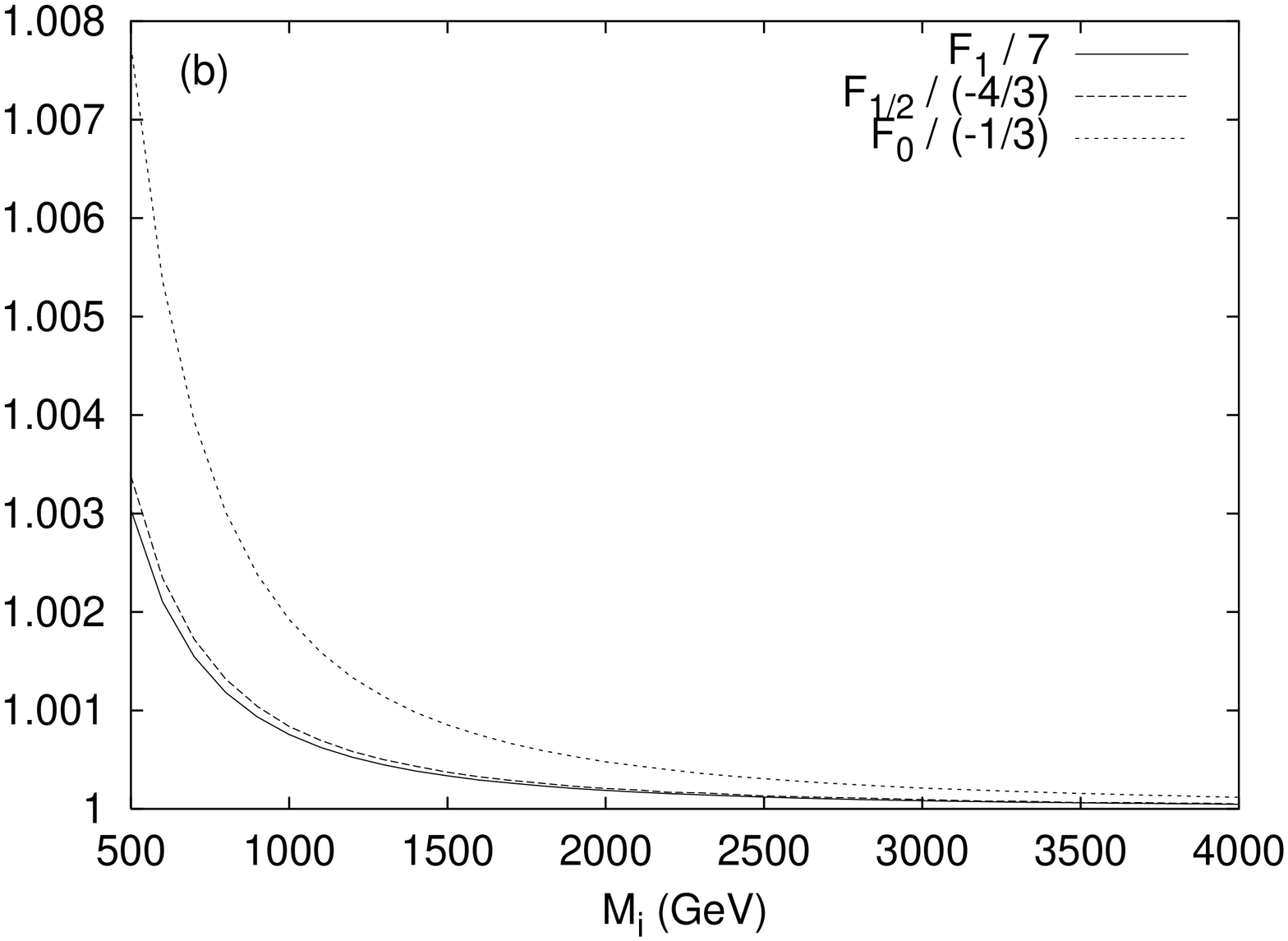}}}
    \caption{(a) The loop functions $F_1(\tau_{W_L})$ and
        $F_{1/2}(\tau_t)$ as a function of $\mh$; (b) The loop functions
        as a function of the heavy mass $M_i$, for $\mh = 120$ GeV.  Both
        figures are normalized to their asymptotic values given in
        Eq.~(\ref{Fasympt}).}
    \label{fig:Fs}
\end{figure}

The partial widths for $H \to gg$ and $H \to \gamma \gamma$ in the SM can 
be found in \cite{HHG}.
In general the partial widths can be expressed as
\begin{equation}
    \begin{array}{lllll}
	\Gamma(H \to gg) & = & \displaystyle
            \frac{\alpha_s^2 m_H^3}{32 \pi^3 v^2}
            \left| \sum_i -\frac{1}{2} y_i F_{1/2}(\tau_i) \right|^2
	&=& \displaystyle
            \frac{\sqrt{2} G_F \alpha_s^2 m_H^3 y^2_{G_F}}{32 \pi^3}
            \left| \sum_i -\frac{1}{2} y_i F_{1/2}(\tau_i) \right|^2, \\
\smallskip
	\Gamma(H \to \gamma\gamma) & = &  \displaystyle
            \frac{\alpha^2 m_H^3}{256 \pi^3 v^2}
            \left| \sum_i y_i N_{ci} Q_i^2 F_i \right|^2
	&=& \displaystyle 
            \frac{\sqrt{2} G_F \alpha^2 m_H^3 y^2_{G_F}}{256 \pi^3}
            \left| \sum_i y_i N_{ci} Q_i^2 F_i \right|^2,
    \end{array}
\label{widths}
\end{equation}
where $N_{ci},\ Q_i$ are the color factor and electric charge respectively
for a particle $i$ running in the loop. 
The dimensionless loop factors for particles of spin given in the 
subscript are~\cite{HHG}
\begin{eqnarray}
	F_1 = 2 + 3 \tau + 3\tau (2-\tau) f(\tau),\quad 
	F_{1/2} = -2\tau [1 + (1-\tau)f(\tau)],\quad
	F_0 = \tau [1 - \tau f(\tau)],
\end{eqnarray}
with
\begin{equation}
	f(\tau) = \left\{ \begin{array}{lr}
		[\sin^{-1}(1/\sqrt{\tau})]^2, & \tau \geq 1 \\
		-\frac{1}{4} [\ln(\eta_+/\eta_-) - i \pi]^2, & \, \tau < 1
		\end{array}  \right.
\end{equation}
and
\begin{equation}
	\tau_i = 4 M_i^2 / m_H^2, \qquad
	\eta_{\pm} = 1 \pm \sqrt{1-\tau}.
\end{equation}
In the limit of large $\tau$, {\it i.e.}, when the particle in the loop is
much heavier than $H$, the loop factors approach constant values:
\begin{equation}
\label{Fasympt}
	F_1 \to 7, \qquad F_{1/2} \to -4/3, \qquad F_0 \to -1/3.
\end{equation}
On the other hand, for $\tau<1$ the loop factor develops an imaginary part
after crossing the real production threshold of particle pairs in the loop.
The loop functions normalized to their asymptotic values are shown in 
Fig.~\ref{fig:Fs}.   
$F_1(\tau_{W_L})$ and $F_{1/2}(\tau_t)$ are presented as a function of $\mh$ 
in Fig.~\ref{fig:Fs}(a). They are just the SM results for the $W$ 
and top-quark
loops. The loop functions $F_1$, $F_{1/2}$ and $F_0$ for the heavy particles 
are shown in Fig.~\ref{fig:Fs}(b) versus the heavy particle mass.
One can clearly see that the loop functions for the heavy particles 
are very close to their asymptotic values.

In Eq.~(\ref{widths}), the coefficients $y_i$ are the correction 
factors of the Higgs boson couplings with respect to the SM values. 
Expressing the resulting loop integrals as the dimensionless functions 
$F_1,F_{1/2},F_0$,
the fact that the SM couplings $y_t,\ywl$ are of order unity is because 
the Higgs boson couplings to $t$ and $W_L$
are proportional to their masses.
This is not true for the heavy particles in the LH model, since they acquire 
their mass not from their coupling to $H$ but rather from the $f$ condensate.  
Consequently, all the corrections due to heavy particles in the loop 
are proportional to $v^2/f^2$, as is evident
from Eq.~(\ref{yi}). This behavior naturally respects the decoupling limit for
physics with much higher scale $f$.

\subsection{$H \to gg$}

\begin{figure}[tb]
    \resizebox{\textwidth}{!}{
        \rotatebox{270}{\includegraphics{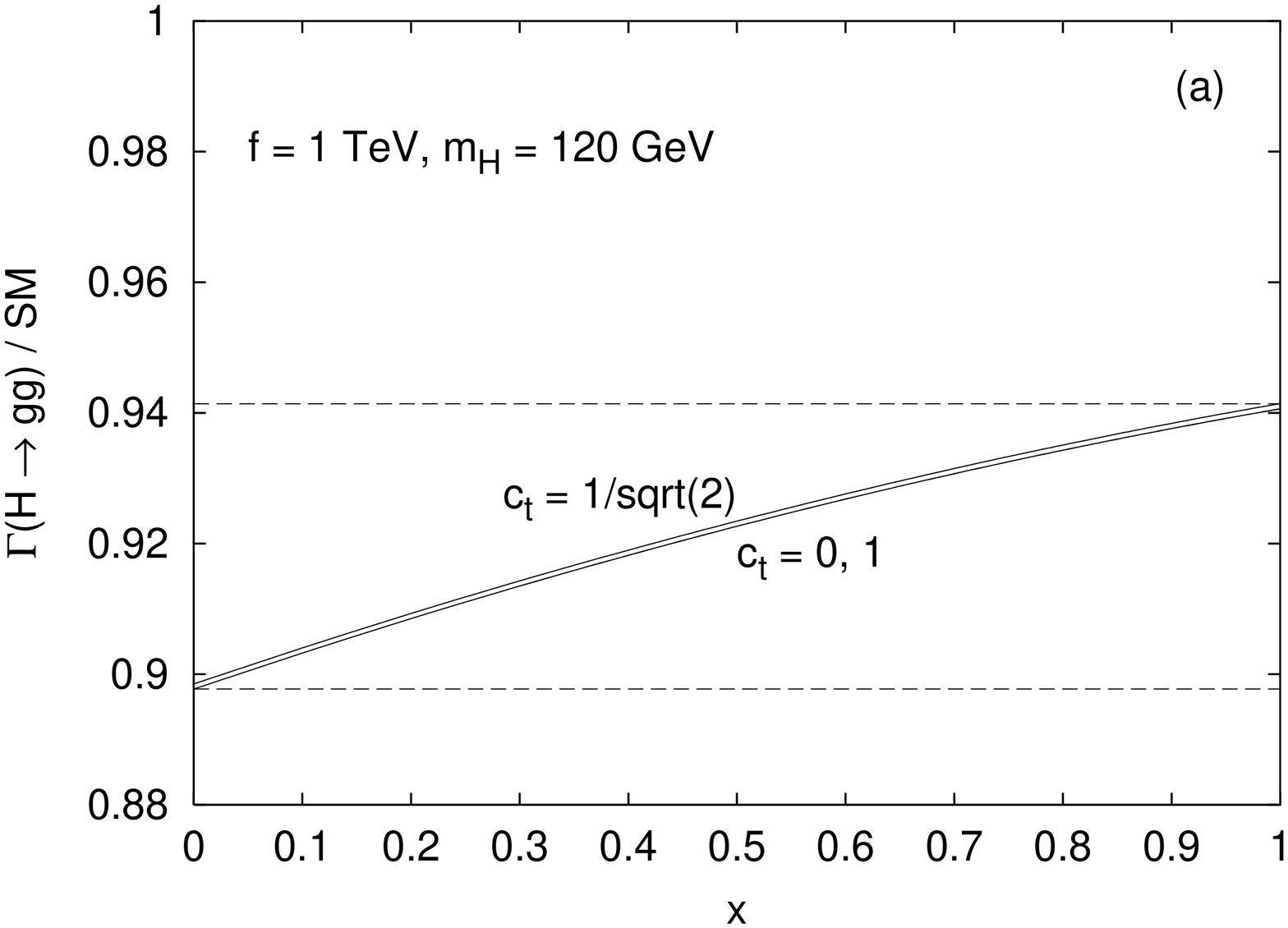}}
        \rotatebox{270}{\includegraphics{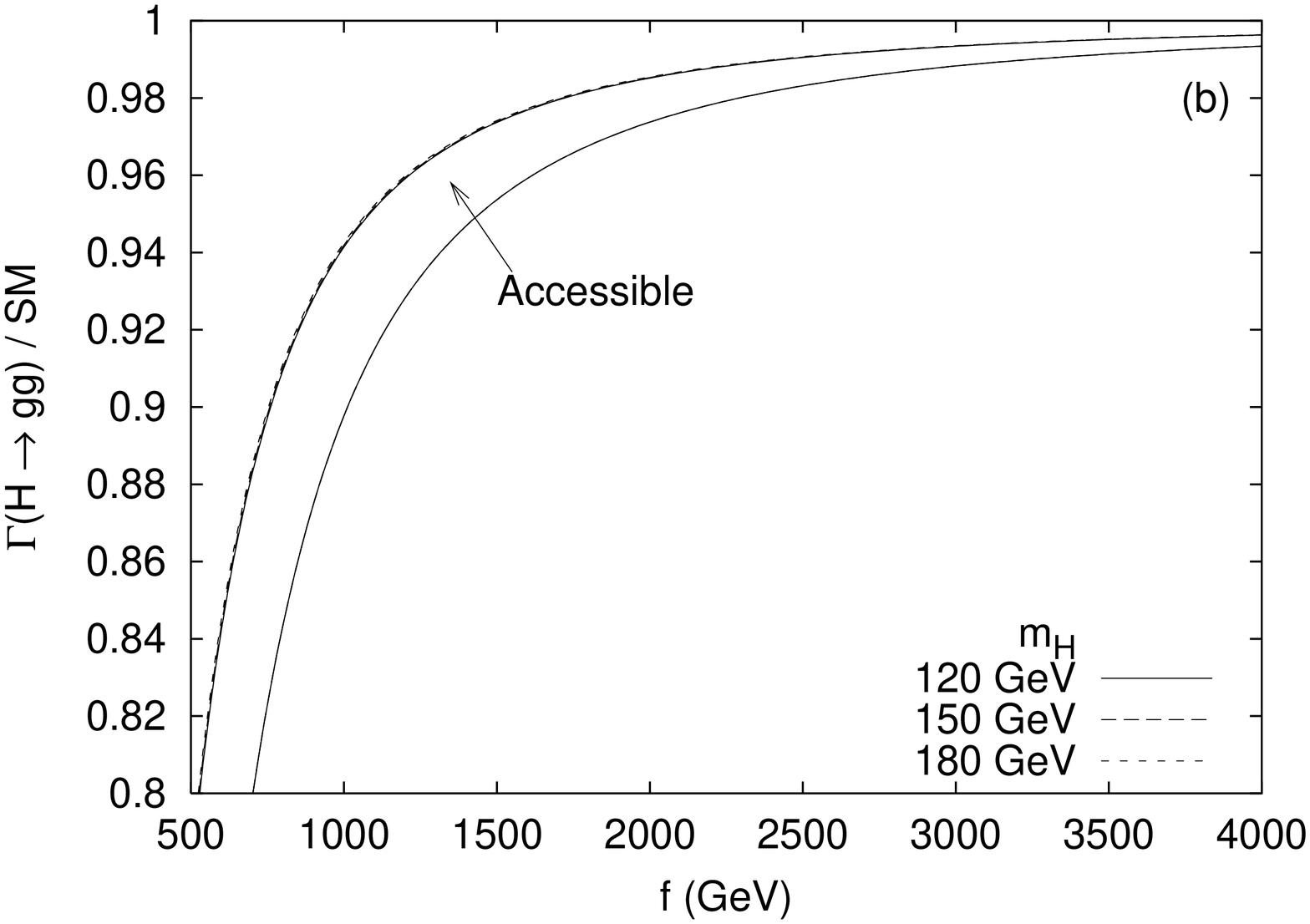}}}
 \caption{(a) Dependence of $\Gamma(H \to gg)$ on the parameters $x$ and
        $c_t$ for $f = 1$ TeV and $\mh = 120$ GeV, normalized to the SM partial width.  
  The solid lines show
        $\Gamma(H \to gg)/{\rm SM}$ as a function of $x$ for $c_t = 1$ or 0
	and $1/\sqrt{2}$ (top to bottom).  The dashed lines indicate the
        minimum ($c_t = 0$ or 1, $x = 0$) and maximum ($c_t = 1/\sqrt{2}$, 
	$x = 1$) values of
        $\Gamma(H \to gg)/{\rm SM}$ obtainable in the LH model for  
        $f=1$ TeV.  (b)  Accessible range of $\Gamma(H \to gg)/{\rm SM}$ 
in the LH model versus $f$ for various values of $\mh$ as indicated.}
    \label{fig:PWgscan}
\end{figure}

In the LH model, the decay $H \to gg$ receives a contribution from  the new
heavy quark $T$. The partial width of $H \to gg$ is given explicitly by
\begin{eqnarray}
	\Gamma(H \to gg) &=& \frac{\sqrt{2} G_F \alpha_s^2 m_H^3}{32 \pi^3}
	\left| -\frac{1}{2} F_{1/2}(\tau_t) y_t y_{G_F} 
	- \frac{1}{2} F_{1/2}(\tau_T) y_T \right|^2
	\nonumber \\
	&=&
	\frac{\sqrt{2} G_F \alpha_s^2 m_H^3}{32 \pi^3}
	\left| -\frac{1}{2} F_{1/2}(\tau_t)
	- \frac{1}{2} \frac{v^2}{f^2} \left[
	\left( -\frac{7}{8} + \frac{1}{2} x - \frac{1}{8} x^2 \right)
	F_{1/2} (\tau_t)
	\right. \right. \nonumber \\
	&& \qquad \qquad \qquad \left. \left. \frac{}{}
	+ c_t^2 s_t^2 \left( F_{1/2} (\tau_t) - F_{1/2} (\tau_T) \right)
	\right] \right|^2.
	\label{eq:Hggbreakdown}
\end{eqnarray}
The first term in the absolute square is the SM result due to the top quark.
Examining the correction terms in the square brackets, we see that
the first term is negative for $0 \leq x < 1$,
leading to a suppression of $\Gamma(H \to gg)$ compared to its SM 
value.  This term comes from $y_{G_F}$ and from the terms in $y_t$
independent of $c_t$ [see Eq.~(\ref{yi})], which are due to the 
mixing in the Higgs sector and the effects of expanding out the
fields of the nonlinear $\sigma$-model in the fermion mass terms.
The second term in the square brackets, which arises from the 
$T$ quark loop plus the mixing
between the SM top quark and the new heavy vector-like quark, 
is positive but too small
to counteract the negative first term.  Note that in the 
limit that $m_t \to M_T$, $F_{1/2}(\tau_t) - F_{1/2}(\tau_T) \to 0$ 
and this second term vanishes.  This is because 
$H$ does not couple to the heavy vector-like quark in the gauge 
eigenstate basis, so the mixing becomes irrelevant as the masses 
become degenerate.
Because $F_{1/2}$ approaches a constant at large $\tau$ and thus 
does not depend
sensitively on $f$, the deviation of $\Gamma(H \to gg)$ from its SM
value simply scales like $1/f^2$.

The dependence of $\Gamma(H \to gg)$ on $x$ is shown in 
Fig.~\ref{fig:PWgscan}(a)
for various values of $c_t$, normalized to the SM partial width.  
We first note that the LH corrections are always
negative, mainly due to the reduced coupling strength of $y_t,\ \ygf$.
There is both linear and quadratic dependence on $x$, but the linear term has a
larger coefficient by a factor of 4, as is evident from the solid curves 
in Fig.~\ref{fig:PWgscan}(a). 
The $x$-dependence leads to a reduction of the partial width varying 
between about $6-10\%$ of the SM value for $f = 1$ TeV, as indicated 
by the range between the dashed lines. 
The $c_t$-dependence of the cross section, as shown by the range of 
the solid curves in the
figure, is quite weak, leading to a variation of the partial width by 
less than 1\%
of the SM value for $f = 1$ TeV.  This is due to the near-cancellation
of the $c_t$-dependence between the $t$ and $T$ loops, as illustrated in
Eq.~(\ref{eq:Hggbreakdown}). For instance,  for $\mh = 120$ GeV, 
$F_{1/2}(\tau_t)/2 \simeq -0.686$ and $F_{1/2}(\tau_T)/2 \simeq -2/3$, 
so the term
proportional to $c_t^2 s_t^2$ has a very small coefficient.  
To further explore the
maximum variation to the SM prediction, we show the accessible range of LH
corrections versus $f$ for various $\mh$ values in Fig.~\ref{fig:PWgscan}(b).
The lower limit is
independent of $\mh$ because it is reached when $y_T = 0$.
In this case, the amplitude is proportional to $F_{1/2} (\tau_t)$ only, so
the $\mh$-dependence cancels in the ratio with the SM partial width.
The upper limit depends on $\mh$ since the $T$ loop contributes, and
$F_{1/2}(\tau_T)$ has a different $\mh$-dependence than $F_{1/2}(\tau_t)$.
While the $\mh$-dependence is rather weak, we see that the reduction from 
the SM prediction can be significant. All of these features can be 
explicitly illustrated by 
examining the numerical coefficients after normalizing the partial 
width to its SM value as
\begin{equation}
        \frac{\Gamma(H \to gg)}{\Gamma_{\rm SM}(H \to gg)}
	= 1 + \left[ -0.106 + 0.061 x - 0.015 x^2
	+ 0.003 c_t^2 s_t^2 \right]
	\left( \frac{1 \ {\rm TeV}}{f} \right)^2,
\label{coeffgg}
\end{equation}
where the $\mh$-dependence is only in the coefficient of the 
$c_t^2 s_t^2$ term; here we have chosen $\mh = 120$ GeV.
Examining Eqs.~(\ref{eq:Hggbreakdown}) and (\ref{coeffgg}), 
we see that the partial width reaches its maximum value (minimum deviation from
the SM) at $x = 1$ and $c_t = 1/\sqrt{2}$, and reaches its minimum value 
(maximum deviation from the SM) at $x = 0$ and $c_t = 0$ or 1.

\subsection{$H \to \gamma \gamma$}

\begin{figure}[tb]
    \resizebox{\textwidth}{!}{
        \rotatebox{270}{\includegraphics{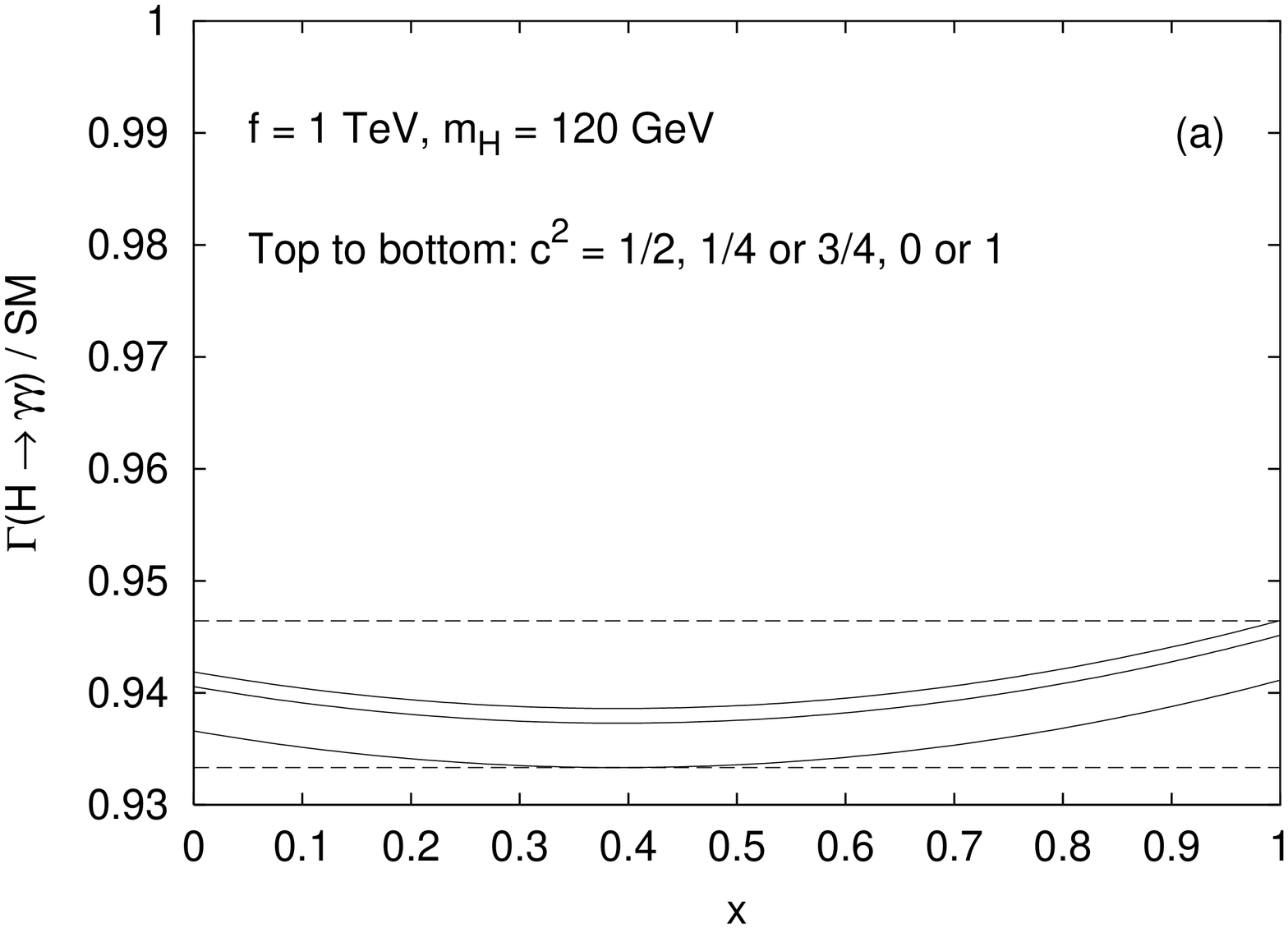}}
        \rotatebox{270}{\includegraphics{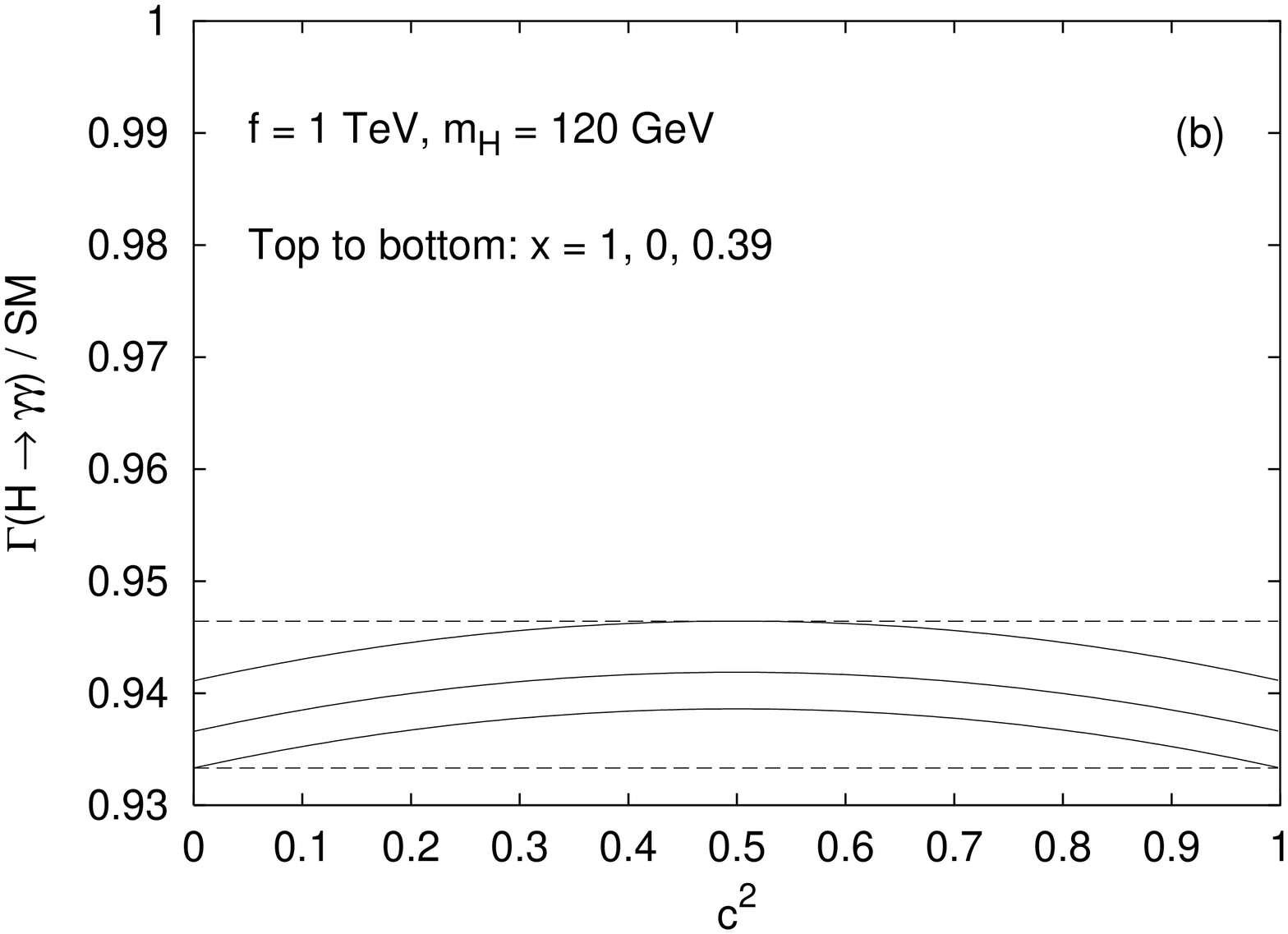}}}
    \caption{Dependence of $\Gamma(H \to \gamma\gamma)$ on the model
        parameters for $f = 1$ TeV and $m_H = 120$ GeV, normalized 
	to the SM partial width. 
        The solid and
        short-dashed lines show $\Gamma(H \to \gamma\gamma)$ relative to
        its SM value as a function of $x^2$ for several
        values of $c^2$ (a) and as a function of $c^2$ for several
        values of $x$ (b).  The solid lines are for $c_t = 0$ and the
        short dashed lines are for $c_t = 1$.  The long dashed lines
        show the minimum ($c=1/\sqrt{2}$, $c_t=x=1$) and maximum ($c=0$
        or 1, $c_t=x=0$) values of $\Gamma(H \to \gamma\gamma)$
        obtainable in the LH model for this value of $f$.  Doubling $f$
        reduces the deviation from 1 by a factor of four.
    }
    \label{fig:PWscan}
\end{figure}

The decay $H \to \gamma\gamma$ in the LH model receives contributions from 
the new
charged particles $W_H^\pm,\ T$, and $\Phi^{\pm}$.
The partial width is given by
\begin{eqnarray}
	\Gamma(H \to \gamma\gamma) &=& 
	\frac{\sqrt{2} G_F \alpha^2 m_H^3}{256 \pi^3}
	\left| \frac{4}{3} F_{1/2} (\tau_t) y_t y_{G_F}
	+ \frac{4}{3} F_{1/2} (\tau_T) y_T
	+ F_1 (\tau_{W_L}) y_{W_L} y_{G_F}  \right. \nonumber \\
	&& \qquad \qquad \qquad \left. \frac{}{}
	+ F_1 (\tau_{W_H}) y_{W_H}
	+ F_0 (\tau_{\Phi}) y_{\Phi^+} \right|^2\nonumber \\ 
         &=&
	\frac{\sqrt{2} G_F \alpha^2 m_H^3}{256 \pi^3}
	\left| 
\frac{4}{3} F_{1/2}(\tau_t) + F_1(\tau_{W_L})
+ \frac{v^2}{f^2} \left[
\frac{4}{3} \left( -\frac{7}{8} + \frac{1}{2} x - \frac{1}{8} x^2 \right)
        F_{1/2}(\tau_t)  \right.   \right.  \nonumber \\ 
        && \quad \left.  \left. \frac{}{} 
        + \frac{4}{3} c_t^2 s_t^2
\left( F_{1/2}(\tau_t) - F_{1/2}(\tau_T)  \right)
+ \left( -\frac{3}{8} - \frac{1}{4} (c^2-s^2)^2 + \frac{1}{8} x^2 \right)
        F_1(\tau_{W_L})  \right.   \right.  \nonumber \\ 
       && \quad \left.  \left. \frac{}{} 
        - s^2c^2 F_1(\tau_{W_H})
        + \left( -\frac{1}{3} + \frac{1}{4} x^2 \right) F_0(\tau_{\Phi}) \right]
	\right|^2.
\end{eqnarray}
The first two terms in the absolute square are the SM results from the 
top quark and $W$ boson. 
For $\mh = 120$ GeV, $4 F_{1/2}(\tau_t)/3 \simeq -1.83$ and 
$F_1(\tau_{W_L}) \simeq 8.17$. In general, for $\mh < 2 M_{W_L}$, the loop
function $F_1$ (for $W_L$ and $W_H$) is real and positive, 
while the loop functions $F_{1/2}$ and $F_0$ (for $t$, $T$, and $\Phi^+$)  
are real and negative. The amplitude is dominated by
the contribution from $W_L$ and is therefore positive.  

The behavior of the contributions of $t$ and $T$ is exactly as in 
the case of $\Gamma(H \to gg)$.  However, because fermion loops
enter with a minus sign in $\Gamma(H \to \gamma\gamma)$, the effects
in the $t$ and $T$ loops that lead to a suppression of 
$\Gamma(H \to gg)$ tend to {\it enhance} $\Gamma(H \to \gamma\gamma)$.
The third term in the square brackets comes from $y_{G_F}$ and $y_{W_L}$
and is always negative.
The term proportional to $(c^2-s^2)^2$ comes from the mixing between
$W_L$ and $W_H$ and suppresses the partial width.
The $W_H$ loop amplitude has a negative coefficient due to the 
$-g$ in the coupling of $W_H^+ W_H^- H$ (see Refs.~\cite{littlesthiggs,us} 
for details), which tends to suppress the partial width.
The last term in the square brackets is due to the $\Phi^+$ loop;
its negative coefficient leads to an enhancement of the amplitude,
but because scalars suffer from the small loop factor of 
$F_0 \simeq -1/3$, its effect is insignificant.
Again, the deviation of $\Gamma(H \to \gamma\gamma)$ from 
its SM value scales like $1/f^2$.

\begin{figure}[h]
    \resizebox{14cm}{!}{
        \rotatebox{270}{\includegraphics{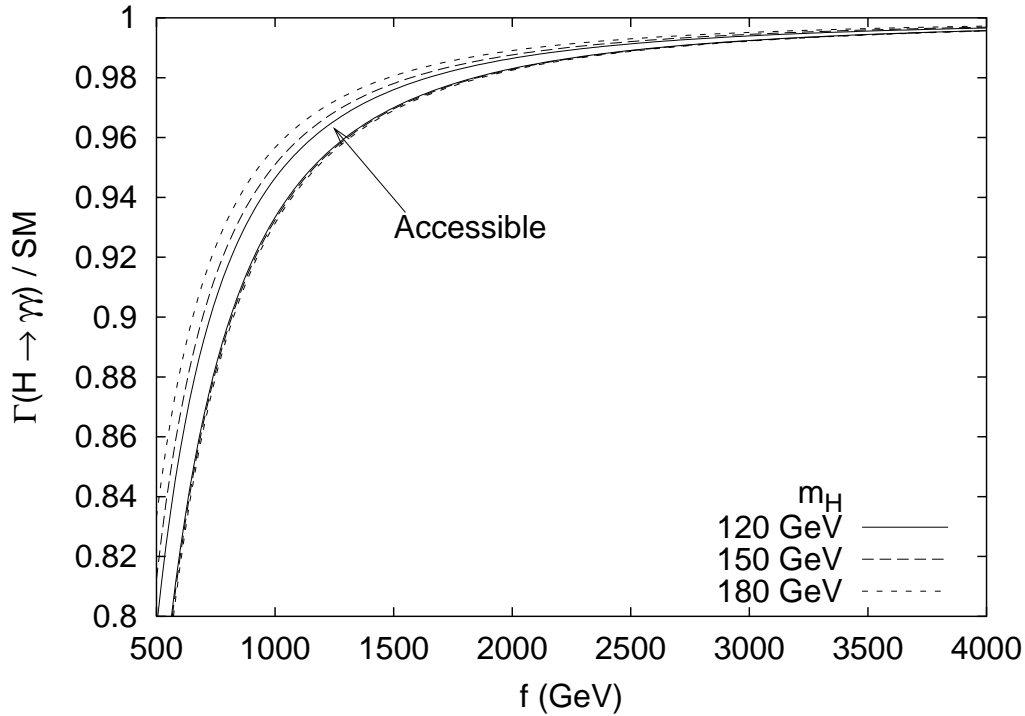}}}
    \caption{Range of values of 
        $\Gamma(H \to \gamma\gamma)$ accessible in the LH model as a
        function of $f$, normalized to the SM value, for $m_H = 120$,
        150 and 180 GeV.}
    \label{fig:range}
\end{figure}

The dependence of $\Gamma(H \to \gamma\gamma)$ on the parameters $c$ and
$x$ is illustrated in Fig.~\ref{fig:PWscan}.  The deviation in
$\Gamma(H \to \gamma\gamma)$ is almost quadratic in $x$ as
seen  in Fig.~\ref{fig:PWscan}(a).  This is due to
the dominant contribution from $F_1(\tau_{W_L})$ with the
$x^2$-dependent coefficients of $\ywl$ and $\ygf$.  Linear
dependence on $x$ enters via $y_t$; however, this contribution is small
because of the small top quark contribution compared to
the $W$ boson contribution.  Varying $x$ between 0 and 1 changes
the partial width by as much as $0.8\%$ of the SM prediction
for $f=1$ TeV.
The partial width is also sensitive to the gauge mixing angle $c$,
due to the $W_L$ and $W_H$ loops. 
It is quadratic in $c^2$, as seen  in Fig.~\ref{fig:PWscan}(b). 
Varying $c$ between $1/\sqrt2$ and 1 or 0 changes the
partial width again by about $0.5\%$ for $f=1$ TeV.
The partial width is almost independent of $c_t$ 
because of the near-cancellation of the $c_t$-dependence between
the $t$ and $T$ loops, as discussed earlier.
The resulting accessible range of $\Gamma(H \to \gamma\gamma)$ is shown in
Fig.~\ref{fig:range}.  
The effect can be quite significant. For instance,
for $f=1$ TeV, the deviation from the SM prediction can be $-(5-7\%)$.
All of these features can be made explicit if we write the expression
normalized to the SM partial width,
\begin{eqnarray}
        \frac{\Gamma(H \to \gamma\gamma)}
        {\Gamma_{\rm SM}(H \to \gamma\gamma)}
	= 1 + \left[ -0.0634 + 0.0211 c^2 s^2
	- 0.0166 x - 0.0211 x^2 - 0.0009 c_t^2 s_t^2 \right]
	\left( \frac{1 \ {\rm TeV}}{f} \right)^2,
\label{coeffgm}
\end{eqnarray}
where we have chosen $\mh = 120$ GeV.
All of the coefficients in the square brackets now depend on $\mh$.

\begin{figure}[tb]
    \resizebox{14cm}{!}{
        \rotatebox{270}{\includegraphics{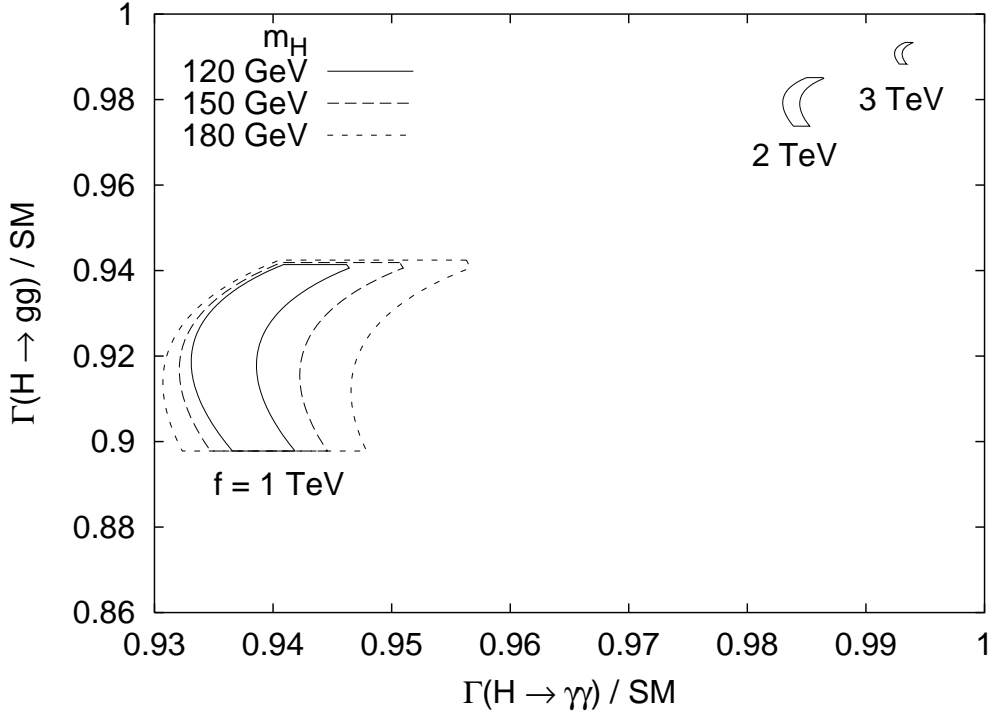}}}
    \caption{Range of values of $\Gamma(H \to gg)$ versus 
$\Gamma(H \to \gamma\gamma)$ accessible in the LH model  normalized to 
the SM value, for $m_H = 120$, 150, 180 GeV and for $f=1,2,3$ TeV.}
    \label{fig:area}
\end{figure}

To summarize the results of the partial width calculations, 
the values of $c$, $c_t$ and
$x$ that minimize and maximize each partial width are given in
Table~\ref{tab:paramextrema}. 
\begin{table}
    \begin{tabular}{|c||c|c|c||c|c|c|}
        \hline
        &\multicolumn{2}{c|}{$\Gamma(H \to gg)$}  &
        Percent &
        \multicolumn{2}{c|}{$\Gamma(H \to \gamma\gamma)$} & Percent \\
        \cline{2-3} \cline{5-6}
        Parameter & Maximize & Minimize & Decrease & Maximize & Minimize & Decrease \\
        \hline
        $c$       & --       & --       & --       & $1/\sqrt{2}$ & 0 or 1 & $0.5\%$ \\
        $x$       & 1        & 0        & $4\%$    & 1        & 0.39            & $0.8\%$ \\
        $c_t$     & $1/\sqrt{2}$ & 0 or 1  & $0.1\%$  & 0 or 1   & $1/\sqrt{2}$ & negligible\\
        \hline
    \end{tabular}
    \caption{Parameters to maximize or minimize the partial widths, and the
    percent change between maximum and minimum (with respect to the SM values)
    for $f=1$ TeV and $m_H = 120$ GeV.  The percent change scales as $1/f^2$.}
    \label{tab:paramextrema}
\end{table}
We have also included the percentage
decrease with respect to the SM predictions when varying the parameters
from the maximum width to the minimum. 
Note that any colored states that enhance 
the $H\to gg$ partial width
would reduce the $H\to \gamma\gamma$ partial width. 
We illustrate the correlation between $\Gamma(H \to \gamma\gamma)$ and
$\Gamma(H \to gg)$ in Fig.~\ref{fig:area}, where the accessible
ranges for $f=1,2,3$ TeV are presented. Also shown is the dependence
on $\mh$ for the case of $f=1$ TeV.

\subsection{Measurements of Higgs couplings to $gg$ and $\gamma\gamma$}

\subsubsection{Higgs couplings at the LHC}

The coupling of $H$ to $gg$ can be probed at the LHC through the
cross section for Higgs production via gluon fusion.  Taking the
ratio of Higgs production rates from gluon fusion and from weak
boson fusion with decays to common final states yields the ratio
of partial widths $\Gamma(H \to gg)/\Gamma(H \to WW)$; the LHC
can measure this ratio with a precision of $25-30\%$ for 
100 GeV $< \mh < 200$ GeV \cite{Dieter}.  The partial width
$\Gamma(H \to gg)$ can be extracted also with a precision of 
$25-30\%$ over this same range of $m_H$ \cite {Dieter}.
The SM partial decay width for $H\to gg$ has been computed accurately to the 
order of $\alpha_s^4$ \cite{widthgg},
and the production cross section for $gg \to H$ has been computed at 
next-to-next-to-leading order \cite{QCDggh}.  The 
remaining renormalization and factorization scale dependence of
the cross section gives a lower bound on the theoretical
uncertainty due to uncomputed higher order QCD corrections of about $15\%$.
This large experimental and theoretical uncertainty may preclude 
the detection of the deviation of the LH model from the SM.

The coupling of $H$ to $\gamma\gamma$ can be probed at the LHC through
$H \to \gamma\gamma$, with $H$ produced via gluon fusion or weak boson
fusion. 
The rates for both $gg \to H \to \gamma\gamma$ and 
$VV \to H \to \gamma\gamma$ can be measured with a precision of 
$10-15\%$ for $m_H < 150$ GeV \cite{Dieter}.
Taking ratios of rates as above but with common Higgs production
mechanisms one can extract the ratio of partial
widths $\Gamma(H \to \gamma\gamma)/\Gamma(H \to WW)$ with a 
precision of $10-20\%$ for 115 GeV $< m_H < 150$ GeV \cite{Dieter}.
The partial width for $\Gamma(H \to \gamma\gamma)$ \cite{widthgmgmh} can be 
extracted with a precision of $15-20\%$ over the same Higgs
mass range \cite{Dieter}.  Such a measurement gives a sensitivity 
only to $f < 650$ GeV at the $1 \sigma$ level,
which is still marginal to reach
the possible new mass scale anticipated in the LH model.

\begin{figure}[tb]
    \resizebox{14cm}{!}{
        \rotatebox{270}{\includegraphics{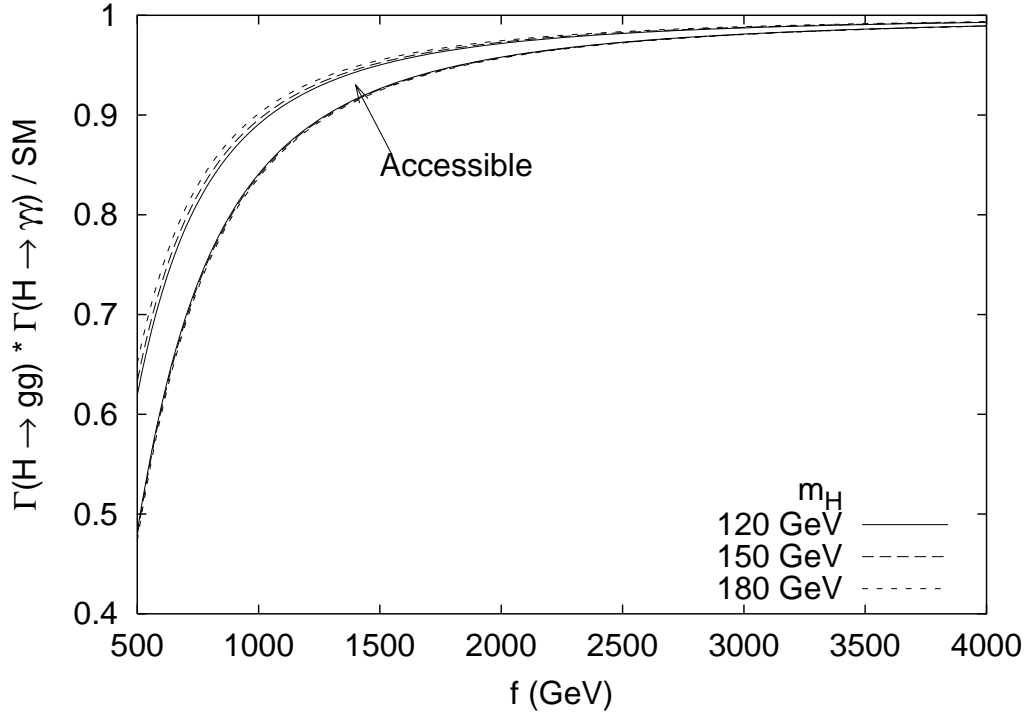}}}
    \caption{Range of values of 
$\Gamma(H \to gg)\times\Gamma(H \to \gamma\gamma)$ 
accessible in the LH model as a
function of $f$, normalized to the SM value, for $m_H = 120$,
        150 and 180 GeV.}
    \label{fig:prodrange}
\end{figure}

As discussed in the previous section, we noticed the correlation 
between $\Gamma(H\to gg)$ and $\Gamma(H\to \gamma\gamma)$.
Since a very promising channel for the Higgs search at the LHC is
$gg\to H\to \gamma\gamma$, it is natural to ask how the product as
a whole is affected. Figure \ref{fig:prodrange} presents the accessible
range for the product 
$\Gamma(H \to gg)\times\Gamma(H \to \gamma\gamma)$  
versus $f$, normalized to the SM value. We see that the effect is to
reduce the production rate with respect to the SM expectation, from $-11\%$ 
to $-16\%$ for $f=1$ TeV.

\subsubsection{Higgs couplings at an $e^+e^-$ collider}

At an $e^+e^-$ collider, the Higgs partial widths can be measured in 
a model-independent way.  The technique involves tagging 
$e^+e^- \to Z^* \to Z H$ events using the $Z$ recoil mass and counting
final states, thus determining absolute branching fractions; 
$e^+e^- \to W^*W^* \to H \nu \bar \nu$ is then used to measure the
$HW^+W^-$ coupling to compute $\Gamma(H \to WW)$ and thus solve for
the individual partial widths.  The cross section for $e^+e^- \to Z H$
is largest at lower center-of-mass energies not too much above the
$ZH$ production threshold.
The cross section for  $e^+e^- \to H \nu \bar \nu$ grows with increasing
center-of-mass energy and can thus be used at higher $\sqrt{s}$ to 
increase the statistics.  At higher energies, however, the $H$ decay
products are more highly boosted, making $b$, $c$ and $\tau$ tagging
more difficult.

The branching ratio of $H$ into $gg$ can be measured with a precision
of about $6-20\%$ at an $e^+e^-$ collider~\cite{LCbook,TESLATDR,ACFA}.  
$H \to gg$ is assumed to be
what is left over after the $b \bar b$, $c \bar c$ and $\tau\tau$ final
states are subtracted off of $H \to$ jets.  This measurement clearly
depends on excellent charm tagging and varies depending on detector
design and machine energy.  As in $gg \to H$ at the LHC,
the theoretical uncertainty in the SM
prediction is likely of order 15\% from uncalculated higher order 
QCD corrections \cite{QCDggh,widthgg}.
Such a measurement would thus be sensitive only to $f < 600$ GeV at the
$1 \sigma$ level.

The measurement of the branching ratio of $H \to \gamma\gamma$ 
at an $e^+e^-$ collider is limited by low statistics due to
the small $H \to \gamma\gamma$ branching ratio $\sim 10^{-3}$.
With 500 fb$^{-1}$ of integrated luminosity one can expect
a precision of about $15-20\%$ on the branching ratio and the 
same for the partial width \cite{LCbook,TESLATDR,Boos}.
This measurement would be sensitive to $f < 650$ GeV at the 
$1 \sigma$ level, comparable
to the sensitivity at the LHC.

\subsubsection{Higgs couplings at a photon collider}

A photon collider can produce the Higgs resonance in the $s$-channel
with a cross section proportional to $\Gamma(H \to \gamma\gamma)$.
For a light Higgs boson with mass around $115-120$ GeV, 
the most precisely measured rate will be
$\gamma\gamma \to H \to b \bar b$, with a precision of about 
2\% \cite{gammagamma}.
(The uncertainty rises to 10\% for $m_H = 160$ GeV.)
This can be combined with the measurement of the branching ratio
of $H \to b \bar b$ measured to about 2\% at the $e^+e^-$ 
collider \cite{LCbook,TESLATDR,ACFA} to extract
$\Gamma(H \to \gamma\gamma)$ with a precision of about 3\%.
Such a measurement would be sensitive to $f < 1500$ GeV at the 
$1 \sigma$ level, or $f < 1100$ GeV at the $2 \sigma$ level.
A $5 \sigma$ deviation is possible for $f < 700$ GeV.

\section{\label{sec:disc}Discussion and Conclusions}

New physics must enter the theory at the cutoff scale 
$\Lambda \simeq 4 \pi f$ to 
complete the non-linear  $\sigma$-model into a linear model.
Because the processes $H \to g g$ and $H \to \gamma\gamma$ are finite
at the leading one-loop order, they do not receive an arbitrarily large
renormalization from cutoff-scale physics.
However, this new physics will generically contain additional 
charged or colored fields that can run in the loop, leading to additional
contributions to $H \to g g$ and $H \to \gamma\gamma$.  
We can estimate their size as follows.  We focus on $H \to \gamma\gamma$,
but similar conclusions can be drawn for $H \to gg$.
The largest contributions will generically come from new gauge bosons,
since their loop factor $F_1$ is the largest in the asymptotic limit.
Consider a single charged gauge boson with mass of order $\Lambda$
and coupling to $H$ of $g_{\Lambda}$.  Its contribution to the amplitude
will be smaller by a factor of  $g^2_{\Lambda}(f^2/\Lambda^2)$
than the contribution from $W_H$, as can be estimated by an argument
of explicit loop calculations or using the Naive Dimensional Analysis~\cite{NDA}.
As long as $g_{\Lambda} < 4 \pi $, the
new physics at the scale $\Lambda$ will not significantly change 
our conclusions, unless there is a large multiplicity of new 
particles that couple to $H$.

While the LH model that we have studied here contains only a single
light Higgs doublet, many little Higgs models in the literature
contain two light Higgs doublets~\cite{littleh}.  The low-energy theory
of these models is therefore a two Higgs doublet model (2HDM), containing
a light charged Higgs boson $H^+$ that can run in the loop
in addition to the new particles at the scale $f$.  Assuming
that the dimensionless Higgs self-coupling $\lambda_{H^+H^-H}$ 
in the 2HDM is of order one as is natural, a relatively light 
$H^+$ with a mass of $100-150$ GeV
can have quite a sizable effect on $\Gamma(H\to\gamma\gamma)$
of $15-40\%$ times $\lambda_{H^+H^-H}$.  For a heavier $H^+$ above 
300 GeV, the deviation is below 5\%.
Potentially more important than the $H^+$ loop, however, is the 
effect of mixing between the neutral components of the two Higgs doublets.
This mixing can affect the couplings of $H$ to $W_L$ and $t$ in a very
significant way if the second Higgs doublet is relatively light, leading
to large deviations in $\Gamma(H\to gg)$ and $\Gamma(H \to \gamma\gamma)$.
If the dimensionless Higgs self-couplings are again of order one,
the deviation of the $HW_L^+W_L^-$ coupling from its SM value 
due to 2HDM mixing is
of order $m_Z^2/m_A^2$ and that of the $Ht\bar t$ coupling 
is of order $m_Z/m_A$,
where $m_A$ is the mass scale of the heavier Higgs doublet (more precisely,
$m_A$ is the mass of the pseudoscalar Higgs boson $A^0$). 
For additional details, see Ref.~\cite{HHG}.
These deviations are thus generically larger than the deviations due to 
the heavy states at the scale $f$.

In conclusion, our results are robust in little Higgs models that contain 
only a single light Higgs boson.  The dominant contributions to
the deviations of $\Gamma(H \to gg)$ and $\Gamma(H \to \gamma\gamma)$
from their SM values are due to (i) the new $T$ quark and $W_H$ bosons in the loop,
which contribute at the order $v^2/f^2$ since they get their masses from 
the condensate $f$; and (ii)
the modification of the $t \bar t H$ and $W_L^+ W_L^- H$ couplings
due to mixing with the heavy states, which also contributes at order
$v^2/f^2$.  Any little Higgs model must contain
such a heavy $T$ and $W_H$ with masses of order $f$
to cancel the quadratic divergence of 
$m_H$ due to the top quark and $W_L$, and these new heavy states
will generically mix with the corresponding SM particles at order
$v^2/f^2$.
Therefore the gross features of our analysis should carry over.

To summarize, we found that for $f = 1$ TeV, $\Gamma(H \to gg)$ 
is reduced by $6-10\%$ in the LH model 
compared to its SM value, where the variation is
mainly due to the dependence on $x$, while $\Gamma(H \to \gamma\gamma)$
is reduced by $5-7\%$ of its SM value, where the variation
is mainly due to the dependence on $x$ and $c$.  The deviations scale
with $1/f^2$.  A photon collider
could probe the deviation in $\Gamma(H \to \gamma\gamma)$ up to
$f \lsim 1.5$ TeV (1.1 TeV, 0.7 TeV) at the $1 \sigma$ 
($2\sigma$, $5\sigma$) level.

\vskip 0.8cm
{\it Note added:}  When the current paper was being completed, 
another paper on a similar  subject appeared \cite{Rogerio}. 
In calculating the Higgs decay widths, the authors of Ref.~\cite{Rogerio} 
did not take into account mixing and interference 
effects between the SM particles and the new heavy states,
and thus reached negligibly small results of the order $(v/f)^4$.

\begin{acknowledgments}
This work was supported in part by the U.S.~Department of Energy under
grant DE-FG02-95ER40896 and in part by the Wisconsin Alumni Research
Foundation.
\end{acknowledgments}

\appendix
\section{}

Here we present some details of the Higgs sector of the LH model
and derive the $H\Phi^+\Phi^-$ and $H\Phi^{++}\Phi^{--}$ couplings.

The most general scalar potential invariant under the Standard Model
gauge groups involving one doublet field $h$ and one triplet field $\phi$
can be written up to operators of dimension four as:
\begin{eqnarray}
	V &=& \lambda_{\phi^2} f^2 {\rm Tr}(\phi^{\dagger} \phi)
	+ i \lambda_{h \phi h} f \left( h \phi^{\dagger} h^T
		- h^* \phi h^{\dagger} \right)
	- \mu^2 h h^{\dagger}
	+ \lambda_{h^4} (h h^{\dagger})^2 \nonumber \\
	&& + \lambda_{h\phi\phi h} h \phi^{\dagger} \phi h^{\dagger}
	+ \lambda_{h^2 \phi^2} h h^{\dagger} {\rm Tr}(\phi^{\dagger} \phi)
	+ \lambda_{\phi^2\phi^2} \left( {\rm Tr}(\phi^{\dagger} \phi) \right)^2
	+ \lambda_{\phi^4} {\rm Tr}(\phi^{\dagger} \phi \phi^{\dagger} \phi).
\end{eqnarray}
The coefficients in
this potential are constrained by the symmetries of the Littlest Higgs model.
At tree-level, there is no Higgs potential.  A Coleman-Weinberg 
potential is generated after the heavy gauge bosons and vector-like quark 
are integrated out.  
The quadratically divergent terms of the one-loop Coleman-Weinberg potential
each preserve one of two global $SU(3)$ symmetries, while breaking the
other (see Ref.~\cite{littlesthiggs} for details).  

From the quadratically divergent part of the one-loop Coleman-Weinberg
potential, we have:
\begin{eqnarray}
    \label{hphipotential}
	\lambda_{\phi^2} &=& \frac{a}{2} \left[ \frac{g^2}{s^2c^2}
	+ \frac{g^{\prime 2}}{s^{\prime 2}c^{\prime 2}} \right]
	+ 8 a^{\prime} \lambda_1^2, \nonumber \\
	\lambda_{h\phi h} &=& -\frac{a}{4} \left[ g^2 \frac{(c^2-s^2)}{s^2c^2}
	+ g^{\prime 2} \frac{(c^{\prime 2}-s^{\prime 2})}
		{s^{\prime 2}c^{\prime 2}} \right]
	+ 4 a^{\prime} \lambda_1^2, \nonumber \\
	\lambda_{h^4} &=& \frac{1}{4} \lambda_{\phi^2}, \quad
	\lambda_{h\phi\phi h} = -\frac{4}{3} \lambda_{\phi^2}, \quad
	\lambda_{\phi^2\phi^2} = -16 a^{\prime} \lambda_1^2, \nonumber \\
	\lambda_{\phi^4} &=& -\frac{2a}{3} \left[ \frac{g^2}{s^2c^2}
	+ \frac{g^{\prime 2}}{s^{\prime 2}c^{\prime 2}} \right]
	+ \frac{16a^{\prime}}{3} \lambda_1^2,
\end{eqnarray}
where $a$ and $a^{\prime}$ are unknown coefficients of order one that
parameterize the effects of the UV-completion at the cutoff scale
$\Lambda$.

The coefficients $\mu^2$ and $\lambda_{h^2\phi^2}$ get no contribution from
the quadratically divergent part of the one-loop Coleman-Weinberg
potential because they are protected by {\it both} of the 
global $SU(3)$ symmetries of the LH model.  Thus they receive only
log-divergent contributions at one-loop, and quadratically divergent
contributions at the two-loop level.
The suppression of $\mu^2$ 
from the extra loop factor gives the natural hierarchy between the 
electroweak scale and the cutoff scale $\Lambda$.  Because
$\lambda_{h^2\phi^2}$ is also generated at this order, it is
not of order one like the other quartic couplings, but rather suppressed
by $1/16\pi^2 \sim 10^{-2}$ and can be neglected for our purposes.

The $H \Phi^+ \Phi^-$ coupling is given by
the following terms in the interaction Lagrangian:
\begin{eqnarray}
	-\mathcal{L} &=& i \sqrt{2} \lambda_{h\phi h} (h^+ h^0 \phi^-
		- h^- h^{0*} \phi^+) f
	+ \frac{1}{2} \lambda_{h\phi\phi h} \phi^+ \phi^- h^0 h^{0*}
	+ \cdots,
\end{eqnarray}
where the dots represent terms that give subleading contributions.
Using the expressions for the couplings~\cite{us},
\begin{eqnarray}
	\lambda_{h\phi h} = \frac{x M_{\Phi}^2}{2 f^2}, \qquad
	\lambda_{h\phi\phi h} 
	= -\frac{4 M_{\Phi}^2}{3 f^2},
\end{eqnarray}
and the mixing angles between the gauge and mass eigenstates given 
in Ref.~\cite{us},
we get for the $H \Phi^+ \Phi^-$ coupling,
\begin{equation}
	-\mathcal{L} = H \Phi^+ \Phi^- 
	\left[ \frac{x^2 v}{2} - \frac{2 v}{3} \right]
	\frac{M_{\Phi}^2}{f^2}.
\end{equation}

The $H \Phi^{++} \Phi^{--}$ coupling is given by
the following terms in the interaction Lagrangian:
\begin{equation}
	-\mathcal{L} = 2 \lambda_{\phi^2\phi^2} \phi^{++} \phi^{--}
		\phi^0 \phi^{0*}
	+ \lambda_{h^2\phi^2} \phi^{++} \phi^{--} h^0 h^{0*}
	+ \cdots,
\end{equation}
where again the dots represent terms that give subleading contributions.
In terms of the mass eigenstates,
\begin{equation}
	-\mathcal{L} = H \Phi^{++} \Phi^{--} \left[ v \times
	\mathcal{O}(v^{\prime 2}/v^2, \, 1/16\pi^2) \right].
\end{equation}
The $H \Phi^{++} \Phi^{--}$ coupling is thus highly suppressed
relative to the $H \Phi^+ \Phi^-$ coupling.


\end{document}